\documentstyle[12pt,graphicx,subfigure]{article}
\def\Journal#1#2#3#4{{#1} {\bf #2}, #3 (#4)}

\def\NPB{{\em Nucl. Phys.} B}
\def\PLB{{\em Phys. Lett.}  B}
\def\PRL{\em Phys. Rev. Lett.}
\def\PRD{{\em Phys. Rev.} D}
\def\ZPC{{\em Z. Phys.} C}

\def\beq{\begin{equation}}
\def\eeq{\end{equation}}
\def\lsim{\ ^<\llap{$_\sim$}\ }
\def\gsim{\ ^>\llap{$_\sim$}\ }
\def\r2{\sqrt 2}
\def\beq{\begin{equation}}
\def\eeq{\end{equation}}
\def\beqn{\begin{eqnarray}}
\def\eeqn{\end{eqnarray}}

\def\sinW2{\sin^2\theta_W}

\def\mz2{M_{z}^2}
\def\c2b{\cos 2\beta}

\def\mz{M_z}

\def\Fq2{F_{2}(q^2)}
\def\sec2w{sec^2\theta_W}
\def\gmin2{(g-2)_\mu}

\catcode`\@=11 % This allows one to modify PLAIN macros.

\def\lsim{\mathrel{\mathpalette\@versim<}}
\def\gsim{\mathrel{\mathpalette\@versim>}}
\def\@versim#1#2{\vcenter{\offinterlineskip
    \ialign{$\m@th#1\hfil##\hfil$\crcr#2\crcr\sim\crcr } }}

\def\PRL{Phys. Rev. Lett.}

\begin{document}
\begin{flushright}
{TIFR/TH/01-41}\\ 
{NUB-TH/3222}\\
%\date{\today}
\end{flushright}
%-----------------------------------
%\documentstyle[preprint,aps]{revtex}
\begin{center}
{\Large\bf $b-\tau$ Unification, $g_{\mu}-2$, the $b\rightarrow s+\gamma$ 
Constraint and  Nonuniversalities\\}
\vglue 0.5cm
{Utpal Chattopadhyay$^{(a)}$ and 
Pran Nath$^{(b)}$
\vglue 0.2cm
{\em 
$^{(a)}$Department of Theoretical Physics,\\ Tata Institute
of Fundamental Research,Homi Bhabha Road\\
Mumbai 400005, India}\\
{\em $^{(b)}$Department of Physics, Northeastern University, Boston,
MA 02115, USA\\} }
\end{center}
\begin{abstract}
An analysis is given of the $b-\tau$ Yukawa coupling unification
in view  of the recent result from Brookhaven on 
$g_{\mu}-2$ and under the constraint of $b\rightarrow s+\gamma$.
We explore $b-\tau$ unification under the above constraints 
for nonuniversal boundary conditions for the soft SUSY breaking
parameters. We find new regions of the parameter space where
significant negative supersymmetric contribution to the b quark
mass can arise and 
$b-\tau$ unification within SU(5) framework can occur with
nonuniversal gaugino masses. 
Specifically we find that for the case 
where the gaugino mass matrix transforms like a 24 plet
one finds a negative contribution to the
b quark mass irrespective of the sign of the Higgs mixing parameter $\mu$
when the supersymmetric contribution to $g_{\mu}-2$ is positive.
We exhibit regions of the parameter space where $b-\tau$
unification occurs for $\mu>0$ satisfying the constraints of $b\rightarrow s+\gamma$ and $g_\mu-2$.  
The $\mu<0$ case is also explored.
The dependence of the accuracy of $b-\tau$ unification defined
by $|\lambda_b-\lambda_{\tau}|/\lambda_{\tau}$ on the parameter space is 
also investigated and it is shown that unification with an accuracy
of a few percent can be achieved in significant regions of the
parameter space. The allowed parameter space is consistent with 
the naturalness constraints and the corresponding sparticle spectrum
is accessible at the Large Hadron Collider.
\end{abstract}

\section{Introduction}
The unification of b and $\tau$ Yukawa couplings, along with  gauge
coupling unification, is traditionally been viewed as a 
success of supersymmetric grand unification models with grand 
unification group structure SU(5) and SO(10).
However, a close scrutiny reveals that $b-\tau$ unification 
is rather sensitively dependent on the parameter space of the 
supersymmetric models such as $\alpha_3$ and 
$\tan\beta$\cite{arason,barger},
 GUT threshold corrections and the effects of gravitational
smearing\cite{das}. 
It is also known that $b-\tau$ 
unification is sensitive to the sign of the Higgs mixing parameter
$\mu$\cite{bagger,deboer} and that it prefers the negative sign of 
$\mu$ in the standard $\mu$ convention\cite{sugra}. 
However, the recent experimental results from
Brookhaven\cite{brown} indicate that for a class of SUSY models the sign of
$\mu$ is positive\cite{chatto2} and this result then  makes the realization of 
$b-\tau$  unification more difficult. 
Some recent analyses have tried to address this
problem. In the analysis of  Ref.\cite{bf}, the authors work in
an SO(10) supersymmetric grand unified model and show that with
the inclusion of the D-term contribution to the sfermion and Higgs
masses arising from the breakdown of SO(10) it is possible to 
obtain Yukawa coupling unification up to about 30\% with a 
positive value of $\mu$. The sparticle spectrum found in this 
analysis is close to that of an inverted hierarchy model\cite{inverted}.
In this scenario the first and the second generation sparticles masses
are typically above a TeV and $a_{\mu}$  (a=(g-2)/2) is on the extreme low side
of the corridor allowed by the BNL experiment. Another analysis 
within SO(10) is of Ref.\cite{bdr} which requires an almost exact 
Yukawa coupling
unification for positive $\mu$.  The $a_{\mu}$  predicted in this
work is also rather small, i.e., in the range $(5-10)\times 10^{-10}$.
The scenarios of both Refs.\cite{bf,bdr} require $b-t-\tau$ unification
and are for large $\tan\beta$, i.e., typically $\sim 50$.

In this paper we carry out an analysis with focus on $b-\tau$ Yukawa 
unification within SU(5). We do a comprehensive
 study of this problem with inclusion of nonuniversalities. 
We find new regions of the parameter space 
where one has $b-\tau$ unification  with 
satisfaction of the $g_{\mu}-2$ and the $b\rightarrow s+\gamma$
constraints. 
While most of the new parameter space where the desired
constraints are satisfied requires $\mu>0$, we also find regions of
the parameter space satisfying all the constraints for the 
$\mu<0$ case. 
The outline of the rest of the paper
is as follows. In Sec.II we discuss briefly the current situation
on $g-2$. In Sec.III we discuss the framework of the proposed
analysis. In Sec.IV we give a discussion of the results. 
Conclusions are given in Sec.V. 

\section{The $g_{\mu}-2$ Constraint on SUSY}
We begin with  a discussion of the current situation on 
$a_{\mu}$. The recent Brookhaven National
Laboratory (BNL) experiment gives\cite{brown} 
\begin{equation}
a_{\mu}^{exp}=11659203(15)\times 10^{-10}
\end{equation}
while the prediction in the Standard  Model where    
$a_{\mu}^{SM}=a_{\mu}^{qed}+a_{\mu}^{EW}+a_{\mu}^{hadronic}$  
is\cite{czar1} 
\begin{equation}
a_{\mu}^{SM}= 11659159.7(6.7)\times 10^{-10}
\end{equation}
Essentially the entire error above arises from the
hadronic correction since\cite{davier,others}
 $a_{\mu}^{had}$ $(vac.pol.)$ $=$ $692.4 (6.2)$
 $\times 10^{-10}$.
There is thus a $2.6 \sigma$ deviation between theory and experiment,

\begin{equation}
a_{\mu}^{exp}- a_{\mu}^{SM}=43(16)\times 10^{-10} . 
\end{equation}
The result above is to be compared with the electro-weak correction
in the Standard Model\cite{czar1} 
$a_{\mu}^{EW}=15.2(0.4)\times 10^{-10}$.
The observed difference between experiment
and theory could arise due to a variety of phenomena\cite{czar2}.
Of special interest to us here is the possibility that the observed
phenomenon is supersymmetry.
 It is well known that SUSY makes important 
contributions to $g_{\mu}-2$\cite{kosower,chatto1}. After the result 
of the BNL experiment became available analyses within SUSY were 
carried out for a  variety of 
scenarios. We begin by summarizing here the results for mSUGRA\cite{chams} 
which is characterized by the following parameters:
$m_0$,  $m_{\frac{1}{2}}$, $A_0$, $\tan\beta$ and $sign(\mu)$  
 where $m_0$ is the universal scalar mass,  $m_{\frac{1}{2}}$
 is the universal gaugino mass, $A_0$ is the universal trilinear
 coupling,  $\tan\beta =<H_2>/<H_1>$ where $H_2$ gives mass to
 the up quark and $H_1$ gives mass to the down quark and the lepton of 
each generation 
and $\mu$ is the Higgs mixing parameter.
In this case one finds that the sparticle spectrum consistent 
 with the observed effect satisfies the constraint so that
 $m_0 \leq 1.5$ TeV and $m_{\frac{1}{2}}<0.8$ TeV and further 
 that the sign of $\mu$ is determined to be positive\cite{chatto2}.
  Now  positive $\mu$ is 
 preferred by the $b \rightarrow  s + \gamma$ constraint
 since  for positive $\mu$, the parameter space consistent with 
 the constraint is large,
 while for negative $\mu$  the parameter space consistent with the 
  constraint is small\cite{bsgamma,bsgammanew}. 
  A positive $\mu$ is also desirable for the direct 
 detection of dark matter\cite{chatto2,ellisdark}.
Within mSUGRA the sparticle spectrum consistent with the 
BNL constraint lies with in reach of the Large Hadron Collider\cite{baerlhc}. 
However, the upper limits are significantly
model dependent\cite{komine,mw,g2review}. Further, there is also a 
significant dependence on the CP violating phases\cite{icn}. 
Such model dependence should be taken account of while interpreting 
the implications of the BNL result.

\section{Framework of Analysis}
The main purpose of this paper concerns the investigation of the
parameter space where $b-\tau$ unification occurs consistent with
the BNL $g-2$ constraint and the $b\rightarrow s+\gamma$ constraint
by relaxing the universal boundary conditions in mSUGRA. 
The universality of the soft parameters in 
mSUGRA model arises from the assumption of a flat Kahler 
potential and the assumption that the gauge kinetic energy function
$f_{\alpha\beta}$ is proportional to $\delta_{\alpha\beta}$ where
$\alpha,\beta =1,..,24$ for SU(5).  
However, the nature of Planck scale physics is still
largely unknown. Thus it is reasonable to investigate more general
scenarios based on curved Kahler potentials and with non gauge singlet
gauge kinetic energy functions. Of course, there 
exist strong constraints on the allowed form of nonuniversalities
from flavor changing neutral currents. In the scalar sector the
flavor changing neutral current constraints still allow for the
presence of significant amounts of nonuniversalities in the 
Higgs doublet sector and in the third generation 
sector\cite{nonuni}. Our focus in this paper is on nonuniversalities
in the gaugino mass sector. In this sector supergravity theories 
in general allow for the presence of an arbitrary gauge kinetic 
energy function  $f_{\alpha\beta}$. In the presence of a curved
Kahler potential this then leads to a gaugino mass matrix 
of the form
$M_{\alpha\beta}$= $\frac{1}{4}$$\bar{e}^{G/2}$$G^a$$(G^{-1})^b_a$$
(\partial f^*_{\alpha\gamma}$$
/\partial z^{*b})f^{-1}_{\gamma\beta}$, 
where $G = -\ln[\kappa^6 W W^*]- \kappa^2 K$. Here, 
$W$ is the superpotential, $K(z,z^*)$ is the Kahler potential,
 $z^a$ are the complex  scalar fields, and  
$\kappa = (8\pi G_N)^{-\frac{1}{2}} = 0.41\times 10^{-18}$ 
GeV$^{-1}$ where $G_N$ is Newton's constant.   Now
$f_{\alpha\beta}$  would have a nontrivial field dependence
in general involving fields which transform as a singlet or a nonsinglet
irreducible representation of the underlying gauge group. 
With imposition of the $SU(3)_C\times SU(2)_L\times U(1)$
gauge invariance the gaugino masses at the GUT scale $M_G (\sim 2 \times 10^{16} ~{\rm GeV})$  
will in general have the form\cite{anderson} 

\beq
 \tilde m_i(0)=m_{\frac{1}{2}}\sum_r c_r n_i^r
\eeq
where $n_i^r$ are characteristic of the representation r and 
 $c_r$ are the relative weights of the various representations.
Since $f_{\alpha\beta}$  transform as  
$({ 24}\times { 24})_{symm}={ 1}+{ 24}+{ 75}
+{ 200}$ the representations r consist of the set 
$\{ 1,24, 75, 200\}$.  The singlet representation
leads to universality of the gaugino masses while the 
nonsinglet representations ${ 24},{ 75}$ and ${ 200}$
 generate nonuniversalities ($n_i^r$ for various 
 representations are given in the Appendix).  
  We will investigate $b-\tau$ unification by
exploring a wide range of the soft SUSY parameter
space with inclusion of nonuniversalities of the type 
that enter in Eq.(4). 
The details of the analysis are as follows:  We begin at the 
GUT scale with a prescribed set of boundary conditions and carry 
out a  two loop renormalization group 
evolution (RGE) for the couplings as well as for the soft parameters.
The electroweak symmetry is broken radiatively by the
minimization of the Higgs potential computed to the 
complete one-loop level~\cite{oneloopeff} at the scale\cite{ccn} 
$Q \sim \sqrt{m_{{\tilde t}_1}m_{{\tilde t}_2}}$.
In the analysis we include the supersymmetric  
 corrections~\cite{susybtmass} 
to the top quark (with $M_t=175$ GeV) and the bottom quark 
mass. For the light Higgs boson mass we have used the code 
{\it FeynHiggsFast}~\cite{feynhiggs}.  

In the analysis we impose the BNL $g_\mu-2$ constraint and the 
$b\rightarrow s+\gamma$ constraint which we now summarize.  
In the Standard Model the branching ratio for the process 
 $b\rightarrow s+\gamma$  is estimated to be\cite{sm} 
 $B(b\rightarrow s+\gamma)$$=$ 
$(3.29\pm  0.33)$$\times 10^{-4}$.
 The most recent experimental determination of this branching ratio
 gives\cite{cleo} 
$B(b\rightarrow s+\gamma)$$=$$ (3.15\pm 0.35\pm 0.32\pm 0.26)$
$\times 10^{-4}$ where the first error is statistical, 
and there are two types of systematic errors. The experimental 
determination of 
$B(b\rightarrow s+\gamma)$ has many inherent errors and so we use  
a $2\sigma$ range around the current experiment\cite{cleo} 
and thus take 
 
\beq
 2\times 10^{-4}<B(b\rightarrow s+\gamma)<4.3 \times 10^{-4}
\eeq
The issue of $b-\tau$ unification is closely tied with the 
supersymmetric correction to the b quark mass 

\beq 
m_b(M_Z)=\lambda_b(M_Z)\frac{v}{\sqrt 2}\cos\beta(1+\Delta_b) 
\eeq
where $\Delta_b$ is loop correction to $m_b$. The running b quark mass
$m_b(M_b)$, where $M_b$ is the pole mass, is computed via  the RG
running from $M_Z$ to $M_b$ using 2-loop
Standard  Model renormalization group equations. The pole mass $M_b$
is then related to the running mass  $m_b(M_b)$ by the 
relation\cite{arason}
\beq 
M_b=(1+\frac{4\alpha_3(M_b)}{3\pi}+12.4\frac{\alpha_3(M_b)^2}{\pi^2})
m_b(M_b)
\eeq
Now the largest contributions to $\Delta_b$ arise from the gluino
 and the chargino pieces\cite{susybtmass}. For the gluino one has 

\beq
\Delta_b^{\tilde g}= \frac{2\alpha_3\mu M_{\tilde g}}
{3\pi}   \tan\beta I(m_{\tilde b_1}^2, m_{\tilde b_2}^2,M_{\tilde g}^2)
\eeq
where 
\beq
I(a,b,c)
=\frac{abln(a/b)+bcln(b/c)+acln(c/a)}{(a-b)(b-c)(a-c)}
\eeq
We note that since $I(m_{\tilde b_1}^2, m_{\tilde b_1}^2,M_{\tilde g}^2)$ 
is always positive one finds that $\Delta_b^{\tilde g}$ is negative when 
 $\mu M_{\tilde g}$ is negative. This situation can
 arise when either $\mu$ is negative and $M_{\tilde g}$ positive or 
 when $\mu$ is positive  and $M_{\tilde g}$ is negative.
 There is a similar situation for the 
 case of the chargino contribution which is 
\beq
\Delta_b^{\tilde \chi^+}= \frac{Y_t\mu A_t}
{4\pi} \tan\beta I(m_{\tilde t_1}^2, m_{\tilde t_2}^2,\mu^2)
\eeq
where $Y_t=\lambda_t^2/4\pi$. 
 In this case $\Delta_b^{\tilde \chi^+}$ is negative when either
 $\mu$ is negative and  $A_t$ is positive or when $\mu$ is positive
 and $A_t$ is negative.  
 We note that typically  satisfaction of the $g_\mu-2$ constraint requires
 that the  sign of $\mu \tilde m_2$ be positive. For universal 
 boundary conditions this would lead to a positive 
 $\Delta_b^{\tilde g}$ which is not preferred by $b-\tau$ unification.
 We note, however, that for the case when the gaugino masses arise
 from a  24 plet term in Eq.(4) one has that the signs of the $\tilde m_3$
and $\tilde m_2$ are necessarily opposite.  Because of this one finds that 
a positive  $a_{\mu}^{SUSY}$ implies a negative 
 $\Delta_b^{\tilde g}$ which is what is preferred by $b-\tau$ unification.
 This anti-correlation between the positivity of $a_{\mu}^{SUSY}$ and the
 negativity of $\Delta_b^{\tilde g}$ occurs {\it {independent of the
 sign of $\mu$}}. However, since the satisfaction of 
 $b\rightarrow s+\gamma$ typically prefers a positive $\mu$ 
 the positive $\mu$ sign would still be preferred but solutions with
 negative $\mu$ are not necessarily  excluded just because of the
 BNL constraint in this case.
 (The fact that a  negative $\mu$  can still give a positive supersymmetric 
contribution to $g_{\mu}-2$ in the 24 plet case was also noticed
in Ref.\cite{mw}). 
Typically one expects the gluino
 contribution of Eq.(8) to be the largest supersymmetric contribution.  
 An enhancement of 
  $\Delta_b$ can occur when both $\Delta_b^{\tilde g}$ and 
 $\Delta_b^{\tilde \chi^+}$ are negative. For $\mu>0$ this occurs
 when $C_{24}<0$ and $A_t<0$ and for $\mu<0$ this occurs when 
 $C_{24}>0$ and $A_t>0$.

\section{Discussion of  Results} 
 We give now details of the numerical analysis with focus on the 
 gaugino mass nonuniversalities 
 as discussed above. We find that the most favorable situation 
 arises for the case when $c_{24}$ is negative 
 while cases with nonvanishing $c_{75}$ and $c_{200}$ 
 do not produce $b-\tau$ unification consistent with $g_\mu-2$ and
 $b\rightarrow s+\gamma$ constraints. It is indeed the $g_\mu-2$ constraint 
 which is not satisfied in the latter two scenarios.
 The result of the analysis
 is given in Figs.1 -10 which we now  discuss in detail.
 In analyzing $b-\tau$ unification it is useful to 
 define the parameter $\delta_{b\tau}$ which prescribes the accuracy
 with which $b-\tau$ unification is achieved where 
 
 \begin{equation}
 \delta_{b\tau}=\frac{|\lambda_b-\lambda_{\tau}|}{\lambda_{\tau}}
 \end{equation}
In Fig.(1a) we plot $\delta_{b\tau}$ vs $\tan\beta$ for the following 
range of parameters: $0<m_0<2000$ GeV, 
$-1000 GeV<c_{24}m_{\frac{1}{2}}<1000 GeV$,
$-6000 GeV <A_0<6000 GeV$ and $\mu>0$. The dotted points satisfy
$b-\tau$  unification at the level shown, the crosses additionally satisfy 
$b\rightarrow s+\gamma$ constraint and the filled circles 
satisfy all the constraints, ie., $b-\tau$ unification,
 $b\rightarrow s+\gamma$  and $g_{\mu}-2$ constraints. 
If one uses a criterion that $b-\tau$ unification be
satisfied to an accuracy of 30\% then Fig.(1a) exhibits that there
are significant regions of the parameter space with $\delta_{b\tau}\leq 0.3$.
However, Fig.(1a) also shows that there are appreciable regions in the
parameter space where  $b-\tau$ unification can be satisfied to 
20\%, 10\% and even less than 5\% level. 
 An interesting phenomenon we see in Fig.(1a) is that the 
values of $\tan\beta$ consistent with $b-\tau$ unification lie
in a wide range depending on the accuracy of the $b-\tau$ unification
that is desired. Thus for a $\delta_{b\tau}\leq 0.3$ one 
finds that the allowed values of $\tan\beta$ can be as low as 
$15-20$ and as high as $40-45$. However, a  $b-\tau$ unification with a few percent accuracy 
would require a value of $\tan\beta$ in the vicinity of 30 or above.
An analysis similar to that of Fig.(1a) but with a plot of $\delta_{b\tau}$ vs
$m_0$ is given in Fig.(1b) where the parameter range is as in Fig.(1a)
and $\tan\beta$ lies in the range $2<\tan\beta<55$. We note that
the bulk of $m_0$ values satisfying all the desired constraints
are such that $m_0\leq 1$TeV. An analysis of $\delta_{b\tau}$ vs 
$C_{24}*m_{\frac{1}{2}}$ is 
given in Fig.(1c). Here one finds  quite remarkably that the allowed
points in the parameter space which satisfy the desired constraints
all appear to lie in a rather narrow strip  
$-200~{\rm GeV}<C_{24}*m_{\frac{1}{2}}<0$, where the narrow strip specifically comes from the $g_\mu-2$ constraint.  Finally in Fig.(1d) a plot of
$\delta_{b\tau}$  vs $A_0$ reveals that the bulk of the allowed
 $A_0$ values consistent with all the desired constraints 
are negative. In most of the parameter space the dominant supersymmetric
correction to the b quark mass arises from the gluino exchange 
term. These results are consistent with our discussion at the
end of Sec.3. 

The importance of supersymmetry for $b-\tau$ unification is exhibited
 in Fig.(2) which gives a plot of $\delta_{b\tau}$
vs $\Delta_b$ where $\Delta_b$ is the supersymmetric contribution 
to the b quark mass. 
 Here the dots refer to the points satisfying $b-\tau$ unification, and filled squares represent points which additionally satisfy 
 the $b\rightarrow s+\gamma$ and the $g_{\mu}-2$ constraints.  
 From Fig.(2) one finds that the
  more accurate the $b-\tau$ unification the larger is the fraction of the 
  supersymmetric contribution
 and that $b-\tau$ unification within a few percent accuracy requires
 a SUSY correction to the b quark mass which can be as large as  
 30\% to 40\%.  Next we discuss $b-\tau$ unification with the stricter
 constraint that the unification hold to within 10\%.  In Fig.(3a) 
 a plot of $\Delta_b$ is given as a function of $\tan\beta$
 with all other parameters the same as in Fig.(2) but with the additional
 constraint that $\delta_{b\tau}\leq 0.1$. Here one finds that
 the SUSY corrections must be at least 15-20\% of the b quark mass but
 could be as large as 40\%. A similar analysis but with 
 $\Delta_b$  vs $m_0$  is given in Fig.(3b), 
 with $\Delta_b$  vs $C_{24} m_{\frac{1}{2}}$
  in Fig.(3c), and  with $\Delta_b$  vs $A_0$ 
  in Fig.(3d). We note that for the case of Fig.(3d) the allowed values of 
  $A_0$ are all negative as anticipated in our discussion at the end of 
  Sec.3 for the $\mu>0$ case. 

\begin{figure}           
\vspace*{-2.0in}                                 
\subfigure[]{                       
\label{perc_params_a} 
\hspace*{-0.6in}                     
\begin{minipage}[b]{0.5\textwidth}                       
\centering
\includegraphics[width=\textwidth,height=\textwidth]{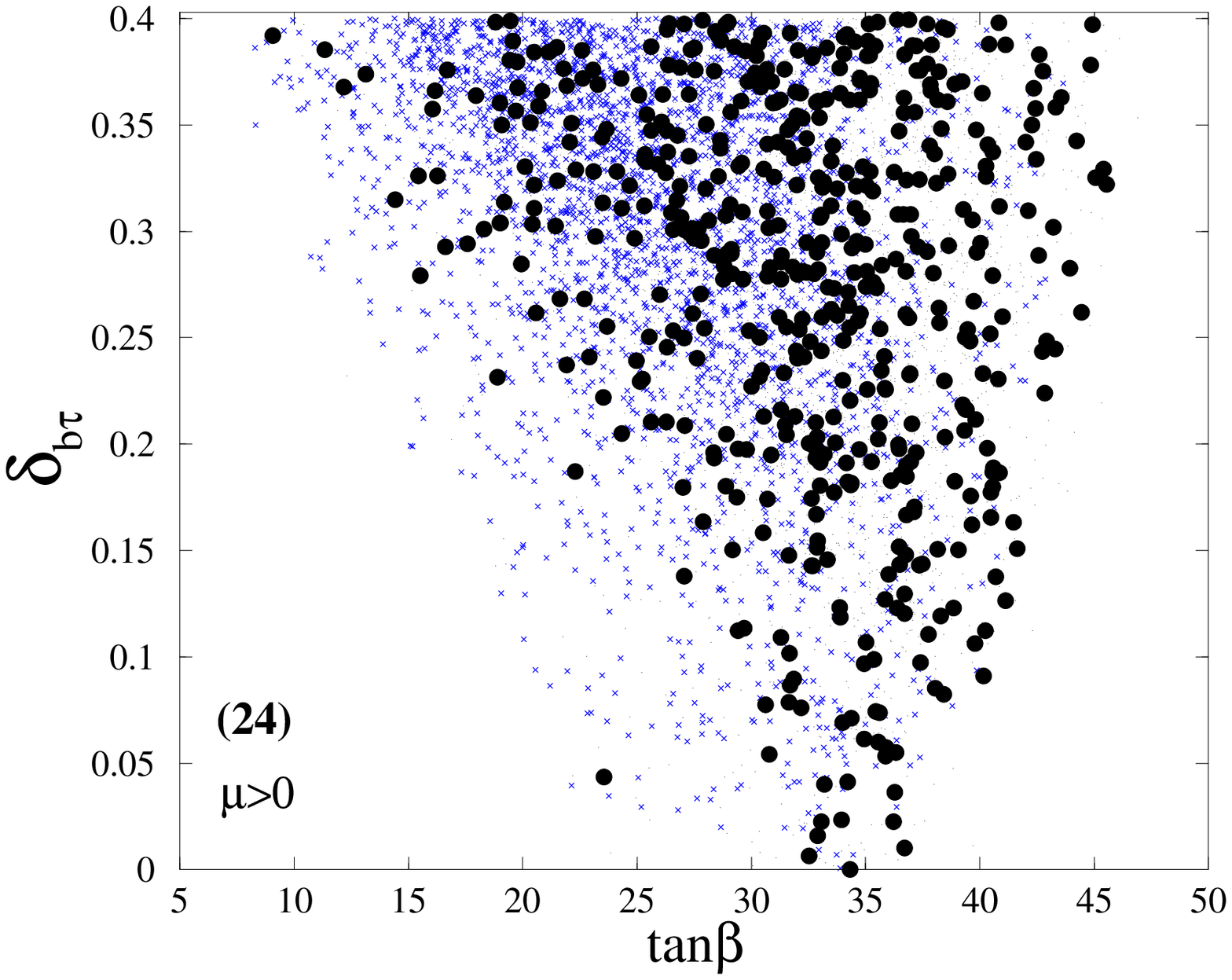}    
\end{minipage}}                       
\hspace*{0.3in}
\subfigure[]{                       
\label{perc_params_b}                       
\begin{minipage}[b]{0.5\textwidth}                       
\centering                      
\includegraphics[width=\textwidth,height=\textwidth]{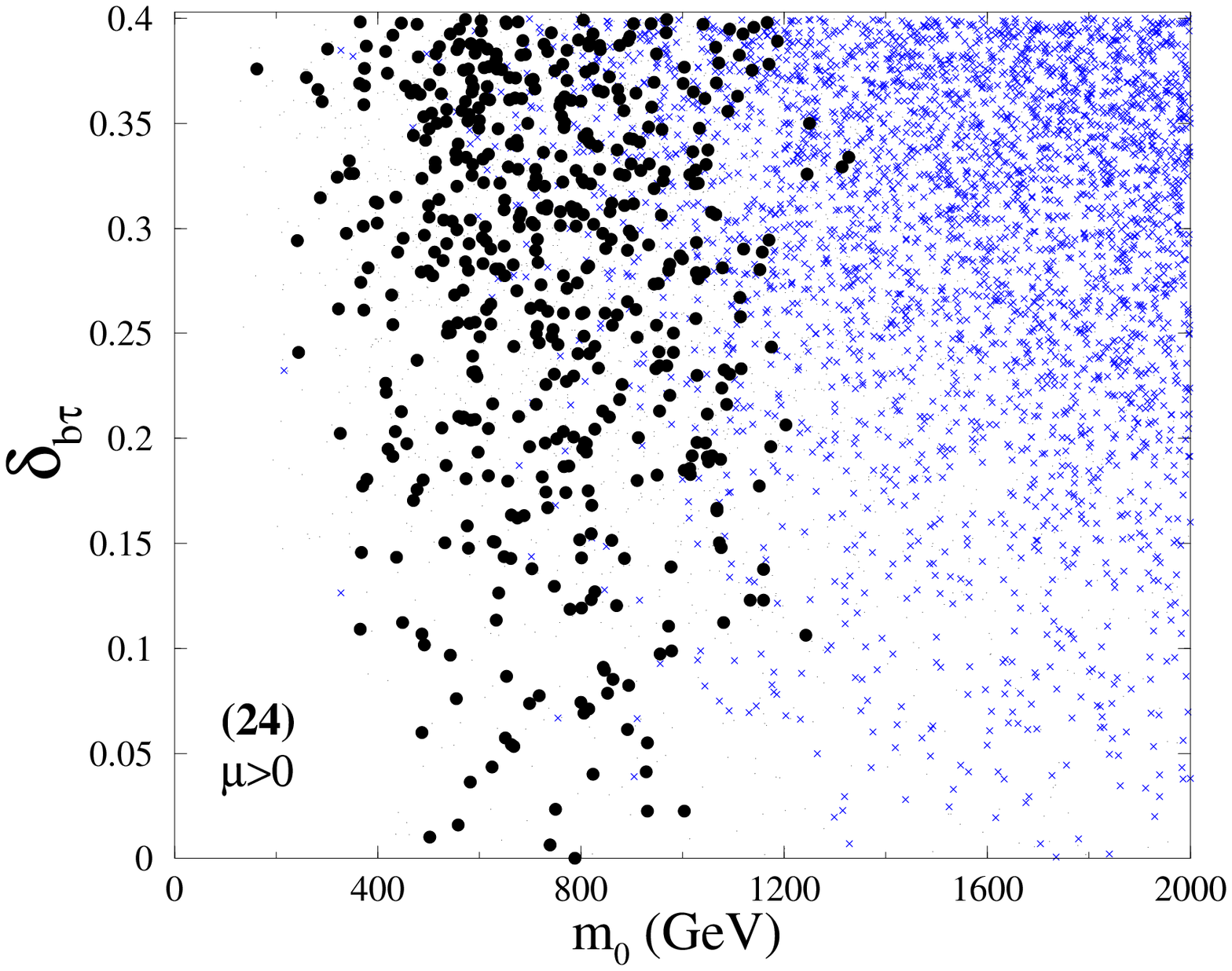} 
\end{minipage}}                       
\hspace*{-0.6in}                     
\subfigure[]{                       
\label{perc_params_c}                      
\begin{minipage}[b]{0.5\textwidth}                       
\centering
\includegraphics[width=\textwidth,height=\textwidth]{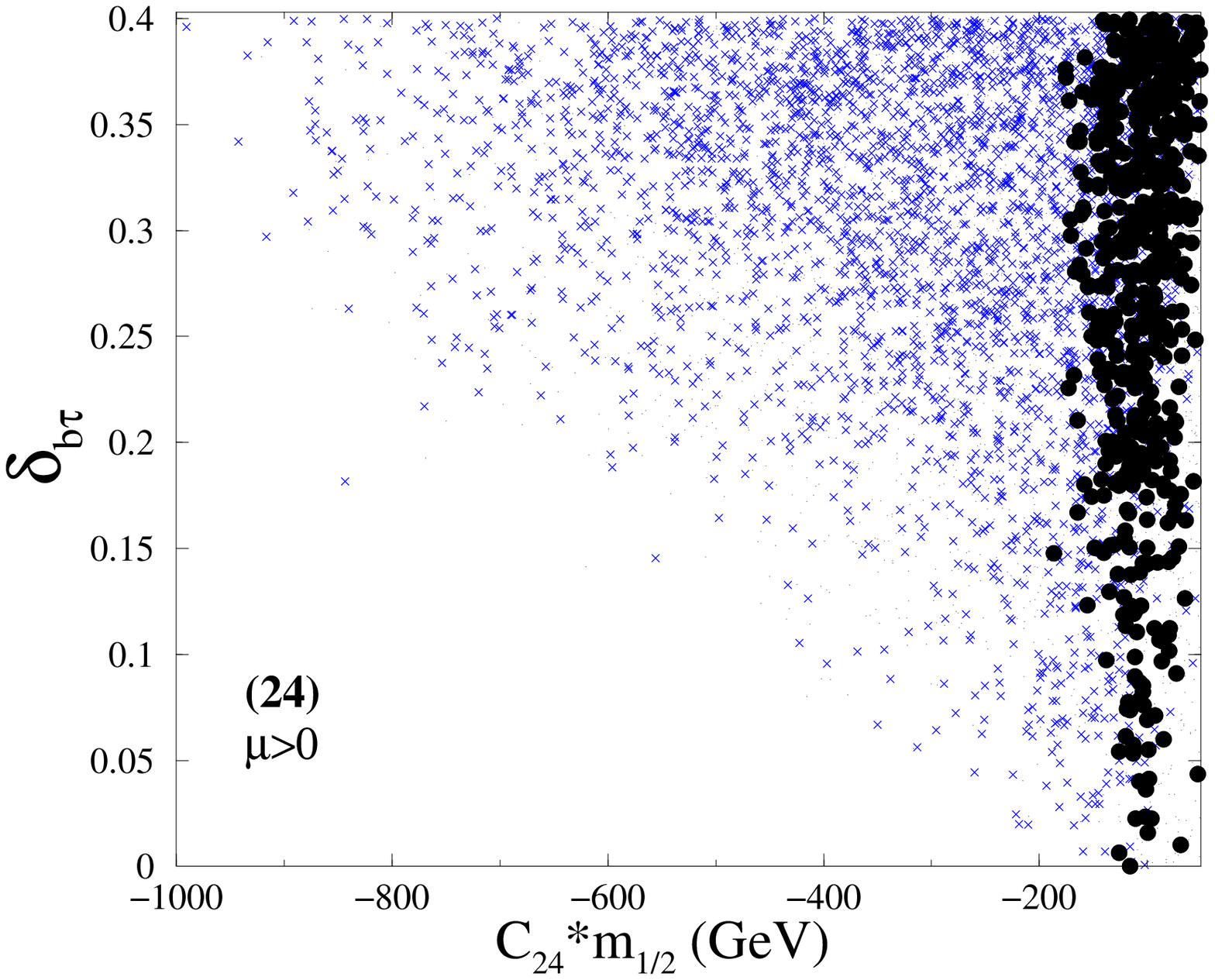}
\end{minipage}}
\hspace*{0.3in}                       
\subfigure[]{                       
\label{perc_params_d}                       
\begin{minipage}[b]{0.5\textwidth}                       
\centering                      
\includegraphics[width=\textwidth,height=\textwidth]{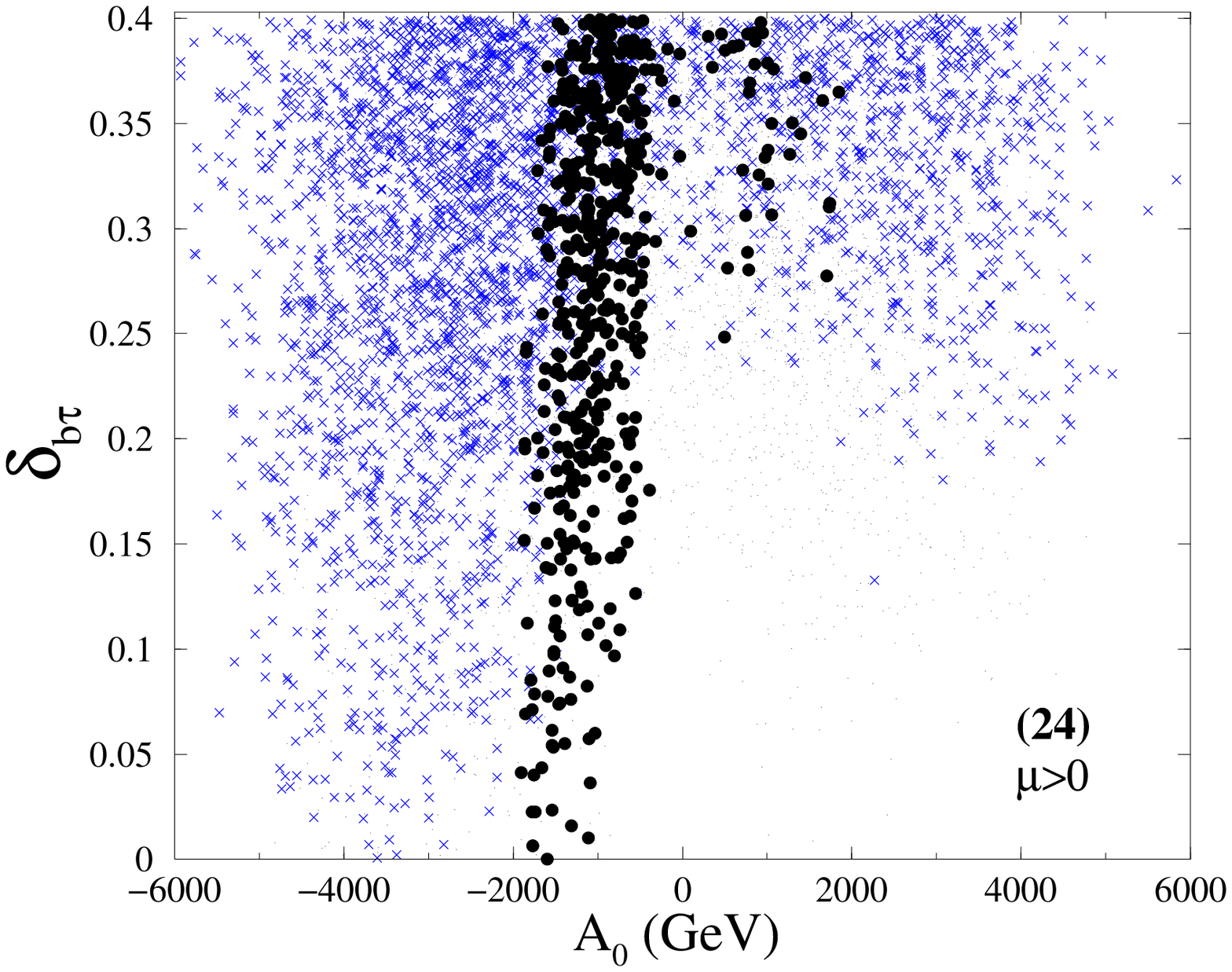}
\end{minipage}}                       
\caption{Plots of $\lambda_b-\lambda_\tau$ unification parameter 
$\delta_{b \tau}$ vs 
model inputs for the 24-plet case when $\tan\beta<55$, 
$0<m_0<2$ TeV, $-1 ~{\rm TeV} <C_{24}m_{1/2}<1 ~{\rm TeV}$,
$-6 ~{\rm TeV}<A_0<6~{\rm TeV}$ and $\mu>0$. The dotted points satisfy
unification at the level shown, (blue) crosses additionally satisfy 
$b \rightarrow s+ \gamma$ limits and (black) filled 
circles satisfy all the constraints, i.e., $b-\tau$ unification at the level
 shown, the $b \rightarrow s+ \gamma$ constraint  and the muon $g-2$ constraint.
}                       
\label{perc_params} 
\end{figure}

Motivated by the fact that $\tan\beta \sim 35$ is the most favored zone
for $b-\tau$ unification (see Fig.~\ref{perc_params_a}) we  compare 
the sparticle mass spectra in mSUGRA with the one arising in the 
24-plet nonuniversal case in 
Figs.\ref{unification_spectra_a} and  \ref{unification_spectra_b}.
Since the $SU(3)$ gauge sector produces important  renormalization group 
effects on the sparticle masses we have chosen identical $m_3(M_G)$ values 
in Figs.\ref{unification_spectra_a} and  \ref{unification_spectra_b} 
for a direct comparison.
Thus for $m_0$ and $m_3(M_G)$ fixed  we find  
that the squark and slepton masses undergo a relatively small change
as we go from Fig.(4a) to Fig.(4b) 
but  a relatively large change is seen 
in the lightest neutralino mass due to the very different ratio of 
the $U(1)$ and $SU(3)$ gaugino masses for the 24 plet case  compared to the 
mSUGRA case. 
  Having  identified  the signature   
of the 24-plet nonuniversal scenario on the mass spectra as given by
Figs.(4a) and (4b) we now study the sparticle 
spectrum in the most favored $b-\tau$ unification region under the
$b \rightarrow s+\gamma$ constraint and the $g_\mu-2$ limits.
The result is exhibited  in 
Fig.\ref{unification_spectra_c}.  The choice of parameters
in Fig.(4c) 
is based on the fact that the requirement of $b-\tau$ unification along with the 
$b \rightarrow s+\gamma$ constraint prefers the parameters 
$\tan\beta=35$, $A_0=-1$~TeV and 
$c_{24}m_{1/2} <0$. The choice 
$c_{24}m_{1/2} =-100$~GeV (see Fig.\ref{perc_params_a}) is guided by
the fact that the $g_\mu-2$ constraint is most easily satisfied for
this value of $c_{24}m_{1/2}$. In Fig.(4c) 
$\delta_{b\tau}$ ranges between $0.2$ to $0.4$. 
Regarding the sparticle spectra we note that as in Fig.(4b) 
the lightest neutralino  mass is much smaller  relative to the 
mSUGRA case of Fig.(4a).
Another interesting phenomenon is that because of the large  $A_0$ 
value in this case there is a large L-R mixing in the stop mixing 
matrix resulting in significantly smaller  $m_{{\tilde t}_1}$ here
relative to that in Fig.(4b). More generally the third generation 
scalar  sector is relatively lighter.
Finally,  most importantly, while considering $m_0$ between 
400 to 800 GeV which is the favored region for satisfying all the constraints 
(see 
 Fig.\ref{perc_params_b}) we see that the spectrum in 
Fig.\ref{unification_spectra_c} 
is well below a TeV. Thus the naturalness requirement is preserved (see, e.g., Ref.\cite{ccn}).

   We discuss now briefly the 24 plet case for $\mu<0$. 
   The results are exhibited in Fig.(5) 
 where we have used the 
 range of parameters $0<m_0<2 TeV$, $-1 TeV <c_{24}m_{1/2}<1TeV$,
 and $-6 TeV<A_0< 6 TeV$ and  imposed 
 the unification criterion $\delta_{b\tau}\leq 0.2$.
 The analysis of Fig.(5) shows that there are significant regions
 of the parameter space  where all the
 constraint, i.e., $b\rightarrow s+\gamma$, $g_{\mu}-2$ and the
 $b-\tau$ unification are consistently satisfied. 
  In Fig.(6) we exhibit the allowed and the disallowed regions due to 
  various constraints in the $m_0-m_3(M_G)$ plane.  
  Dotted points satisfy $b-\tau$ constraint with $\delta_{b\tau}<0.3$, the 
  blue crosses are the points which are additionally allowed by the $g_{\mu}-2$  constraint and the black filled circles satisfy the $g_{\mu}-2$ constraint,
  the $b\rightarrow s+\gamma$ constraint, and the constraint
  $\delta_{b\tau}<0.3$. Fig.(6a) gives the analysis for $\mu>0$ and
  Fig.(6b) for $\mu<0$.
  We note that the density of points where
  all the constraints are satisfied is quite significant for both the
  $\mu>0$ and the $\mu<0$ cases.

 Next we discuss cases where the gaugino masses transform like
 the 75 and the 200 plet representations. We start by 
 exhibiting in Fig.(7) the gaugino masses at the scale $M_Z$  for
 the universal (mSUGRA) case (Fig.(7a)) and for the nonuniversal 
 (24,75 and 100 plet) cases (Figs.(7b-7d)). 
We note that the ratio of gaugino masses for the 75 plet and the 200 plet 
cases are
drastically different from each other and from the mSUGRA and 
the 24 plet cases.
 This is of course what we expect from the nature of the boundary
 conditions for the four cases as given in the Appendix.
In Fig.(8) we exhibit the mass spectra for the universal and the 
nonuniversal cases as a function of $m_0$. One notices the drastic
modification of the pattern of sparticle masses as one goes from 
the universal case of Fig.(8a) to the nonuniversal cases of Figs.(8b-8d). 
A similar 
display  with respect to $m_{\frac{1}{2}}$ for the universal case
and with respect to $m_3(M_G)$ for the nonuniversal cases is given in Fig.(9)
where the gray areas represent the excluded regions because of LEP
chargino mass limits\cite{lep} and the absence of radiative electroweak 
symmetry breaking. Again one finds some drastic modifications of the
sparticle spectrum when one compares the  four cases exhibited in Fig.(9).
In Fig.(10) we exhibit the points in the parameter space 
which satisfy $b-\tau$ unification for 75 and 200 plet cases. 
One finds that there exist significant regions of the parameter space 
in Fig.(10a) and Fig.(10c) 
satisfying $b-\tau$ unification which also additionally satisfy the 
$b\rightarrow s+\gamma$ constraint for $\mu>0$. Both the requirements of 
$b-\tau$ unification as well as the 
$b\rightarrow s+\gamma$ constraint reduces the allowed parameter space
significantly for $\mu<0$, but there
still exist significant regions  of the parameter space in this case which
are allowed by the constraints as shown in Fig.(10b) and Fig.(10d).
However, none of the allowed points in Figs.(10a-10d) 
which satisfy the $b-\tau$ unification 
and the $b\rightarrow s+\gamma$ constraints satisfy the $g_{\mu}-2$ constraint.

\section{Conclusions}
In this paper we have analyzed $b-\tau$ unification within SU(5)
with emphasis on nonuniversalities in the gaugino sector.
We find that there exist regions of the parameter space with 
nonuniversalities in the gaugino mass sector 
which lead to $b-\tau$ unification consistent with
the Brookhaven result on $g_{\mu}-2$ and with satisfaction of the
$b\rightarrow s+ \gamma$ constraint. 
We discussed the 24 plet, the 75 plet and the 200 plet types of 
nonuniversality in the gaugino sector. We found that while $b-\tau$
unification and the $b\rightarrow s+\gamma$ constraint can be satisfied
for all three types of nonuniversalities, but it is only the 
24 plet case where the additional  BNL $g_\mu-2$ constraint can
be satisfied. 
An important result that was observed for the 24 plet case is that
in this case there exists an anti-correlation between the 
positivity of $a^{SUSY}_{\mu}$ and the negativity of $\Delta_b^{\tilde g}$ 
and hence of $\Delta_b$ since $\Delta_b^{\tilde g}$ is the 
dominant contribution to $\Delta_b$ over most of the parameter
space of the model. Quite remarkably the above anti-correlation 
holds irrespective of the sign of $\mu$. The above result implies 
that in the 24 plet case the experimental BNL constraint  that
$a_{\mu}^{SUSY}>0$ automatically leads to a negative contribution 
from $\Delta_b^{\tilde g}$ resulting in a negative $\Delta_b$
which is what is needed for $b-\tau$ unification. 
We investigated both the $\mu>0$ and the $\mu<0$ branches and 
 find satisfactory solutions for both cases. In both cases we find 
 new regions of the parameter space which give negative SUSY contribution
to the b quark mass, satisfy the $g_\mu-2$ as well as 
$b\rightarrow s+\gamma$ constraints and lead to a 
satisfactory $b-\tau$ unification. 
%However, the allowed region for the $\mu>0$ case is significantly larger 
%than for the $\mu<0$ case. 
The new regions of the parameter space are also consistent with the 
naturalness criteria and further the
 corresponding sparticle spectrum  which arises in this parameter space
 lies within the reach of the
 Large Hadron Collider. It would be interesting to discuss the 
 details of the supersymmetric signals such as the trileptonic 
 signal\cite{trilepton,sugra} in this part of the 
 parameter space. However, this analysis is outside the scope of 
 this paper. Further, we have not discussed in this paper the
 topic of proton stability which involves the Higgs triplet sector
 of the theory.   The most recent experimental data from
SuperKamiokande appears to put a rather stringent constraint on 
the Higgs triplet coupling in the simplest supersymmetric grand 
unification models\cite{dmr,mura}. 
However, it is possible to 
relieve this constraint in non minimal  models\cite{altarelli,ns}.
Further work is required in this area but again this analysis 
is outside the scope of this paper. 
While this paper was under preparation the work of Ref.\cite{ky}
appeared where the issue of $b-\tau$ unification with
gaugino mass nonuniversality was also briefly discussed.

\noindent
{\bf Appendix}\\
We record here the numerical values of $n_i^r$ for 
the universal and the nonuniversal cases. For the universal case one has
$n_i^{(1)}=1$ for all i. The cases r=24,75, 200 are nonuniversal.
 For r={24} one has $n_1^{(24)}= -1$; $n_2^{(24)}=-3$; $n_3^{(24)}= 2$.
 For r={ 75} one has $n_1^{(75)}= -5$; $n_2^{(75)}=3$; $n_3^{(75)}= 1$.
Finally, for  r={200} one has  
$n_1^{(200)}= 10$; $n_2^{(200)}=2$; $n_3^{(200)}= 1$.

\noindent
\section{Acknowledgments}
This work was initiated when the authors were at CERN and they
acknowledge the hospitality extended to them during the
period of the visit there. 
This research was supported in part by NSF grant PHY-9901057.

%%%%%%%%%%%%%%%%%%%%%%%%%%%%%%%%%%%%%%%%%%%%%%%%%%%%%%%%
\newpage

%%%%%%%%%%%%%%%%%%%%%%%%%%%%%%%%%%%%%%%%%%%%%%%%%%%%%%%%
\newpage
\begin{figure}
\hspace*{-0.6in}
\centering
\includegraphics[width=\textwidth]{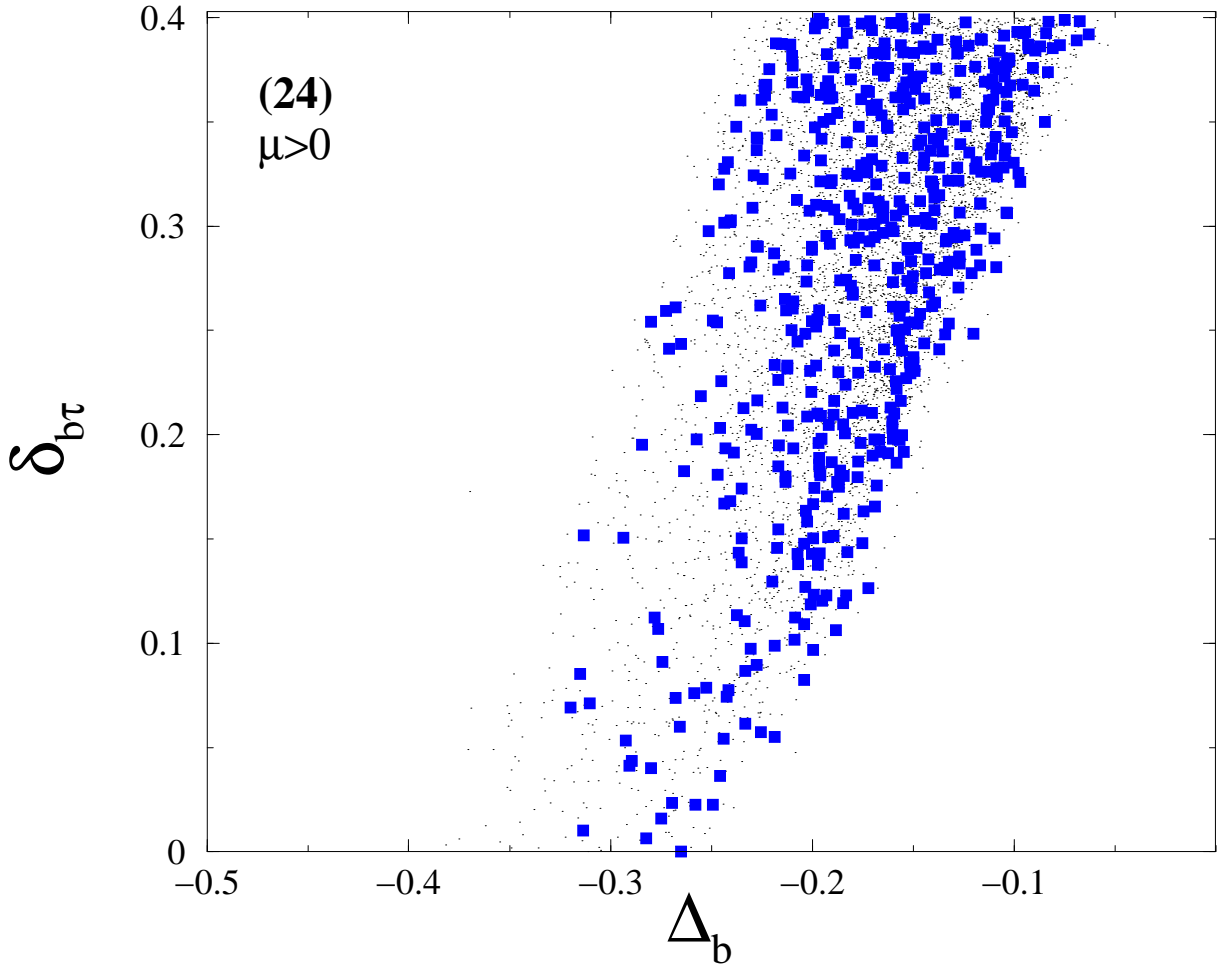}
\caption{Plot of $\lambda_b-\lambda_{\tau}$ unification parameter 
$\delta_{b \tau}$
vs the supersymmetric correction $\Delta_b$ to the b-quark mass for the 
24-plet case, when $\tan\beta<55$,
$0<m_0<2$ TeV, $-1 ~{\rm TeV} <C_{24}m_{1/2}<1 ~{\rm TeV}$,
$-6 ~{\rm TeV}<A_0<6~{\rm TeV}$ and $\mu>0$.
The dots refer to $b-\tau$ unification at the shown level 
and filled (blue) squares additionally
represent points which satisfy both the $b \rightarrow s+\gamma$ and 
the muon $g-2$ constraints.}
\label{perc_botsusy}
\end{figure} 

%%%%%%%%%%%%%%%%%%%%%%%%%%%%%%%%%%%%%%%%%%%%%%%%%%%%%%%%
\newpage
\begin{figure}           
\vspace*{-2.0in}                                 
\subfigure[]{                       
\label{bot_params_a} 
\hspace*{-0.6in}                     
\begin{minipage}[b]{0.5\textwidth}                       
\centering
\includegraphics[width=\textwidth,height=\textwidth]{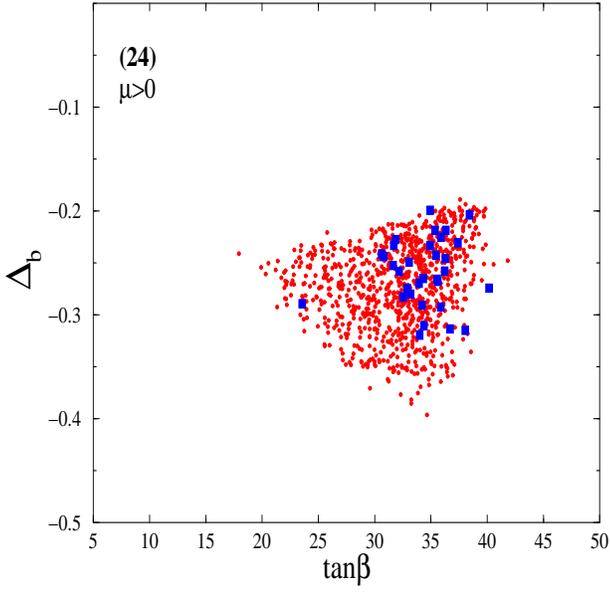}    
\end{minipage}}                       
\hspace*{0.3in}
\subfigure[]{                       
\label{bot_params_b}                        
\begin{minipage}[b]{0.5\textwidth}                       
\centering                      
\includegraphics[width=\textwidth,height=\textwidth]{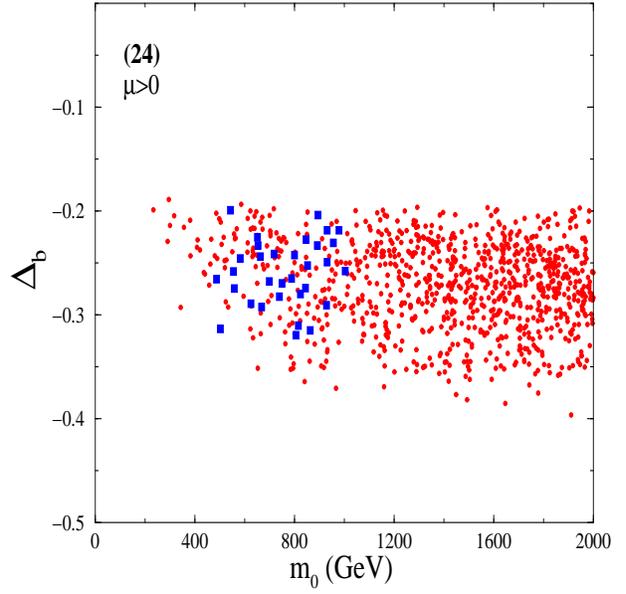} 
\end{minipage}}                       
\hspace*{-0.6in}                     
\subfigure[]{                       
\label{bot_params_c}                       
\begin{minipage}[b]{0.5\textwidth}                       
\centering
\includegraphics[width=\textwidth,height=\textwidth]{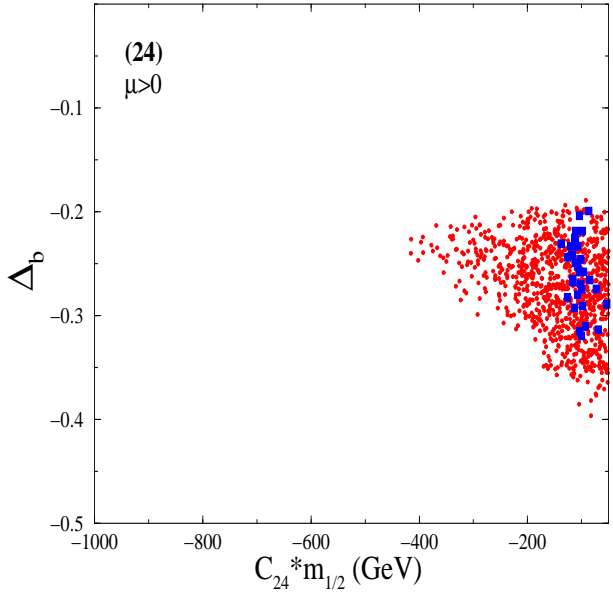}
\end{minipage}}
\hspace*{0.3in}                       
\subfigure[]{                       
\label{bot_params_d}                        
\begin{minipage}[b]{0.5\textwidth}                       
\centering                      
\includegraphics[width=\textwidth,height=\textwidth]{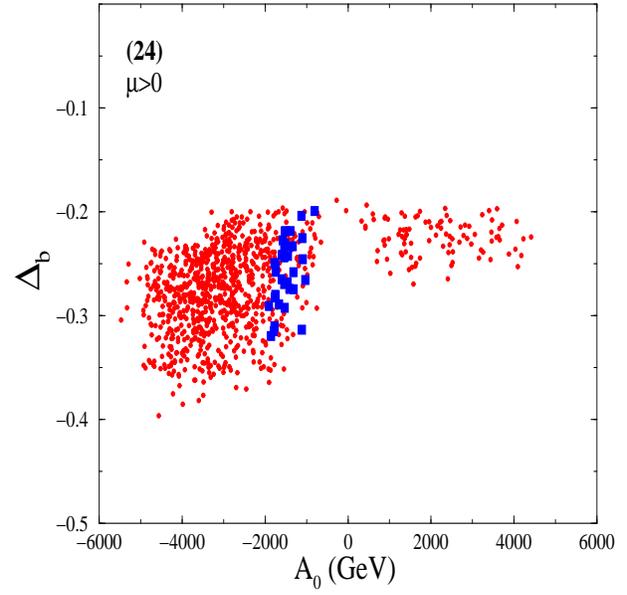}
\end{minipage}}                       
\caption{Plots of the supersymmetric correction $\Delta_b$ to the 
b-quark mass vs model inputs for the 
24-plet case when $\delta_{b \tau}<10$\%.  Here $\tan\beta<55$, 
$0<m_0<2$ TeV, $-1 ~{\rm TeV} <C_{24}m_{1/2}<1 ~{\rm TeV}$,
$-6 ~{\rm TeV}<A_0<6~{\rm TeV}$ and $\mu>0$. 
The (red) dots refer to points satisfying $b-\tau$ 
unification with $\delta_{b \tau}<10$\% and filled (blue) 
squares additionally represent points which satisfy both 
the $b \rightarrow s+\gamma$ and the muon $g-2$ constraints.
}                       
\label{bot_params} 
\end{figure}

%%%%%%%%%%%%%%%%%%%%%%%%%%%%%%%%%%%%%%%%%%%%%%%%%%%%%%%%
\newpage
\begin{figure}                       
\vspace*{-1.0in}                                 
\subfigure[]{
\label{unification_spectra_a}
\hspace*{-0.6in}                     
\begin{minipage}[b]{0.5\textwidth}                       
\centering
\includegraphics[width=\textwidth]{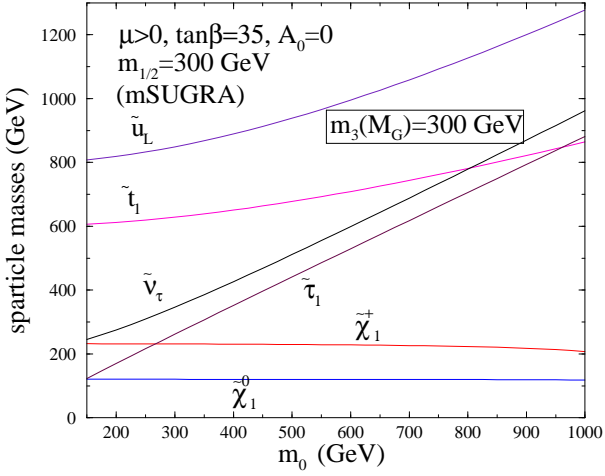}                      
\end{minipage}}
\hspace*{0.3in}                       
\subfigure[]{
\label{unification_spectra_b}    
\begin{minipage}[b]{0.5\textwidth}                       
\centering                      
\includegraphics[width=\textwidth]{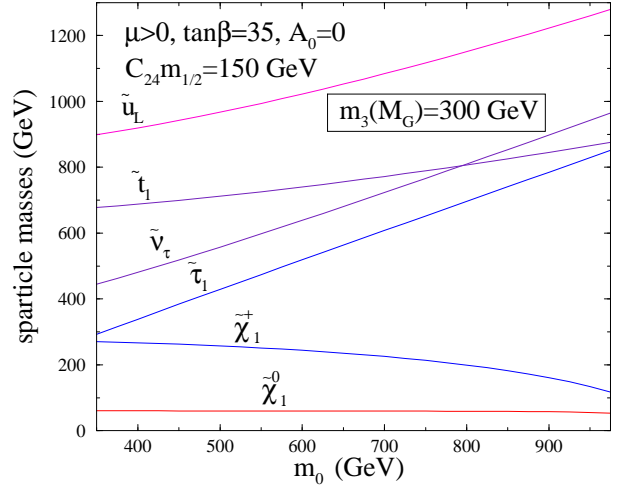}                      
\end{minipage}}                       

%Keep a line blank as above for to have the followign fig in the bottom 
\subfigure[]{
\label{unification_spectra_c}             
\hspace*{-0.4in}                               
\begin{minipage}[b]{\textwidth}                       
\centering                      
\includegraphics[width=0.7\textwidth]{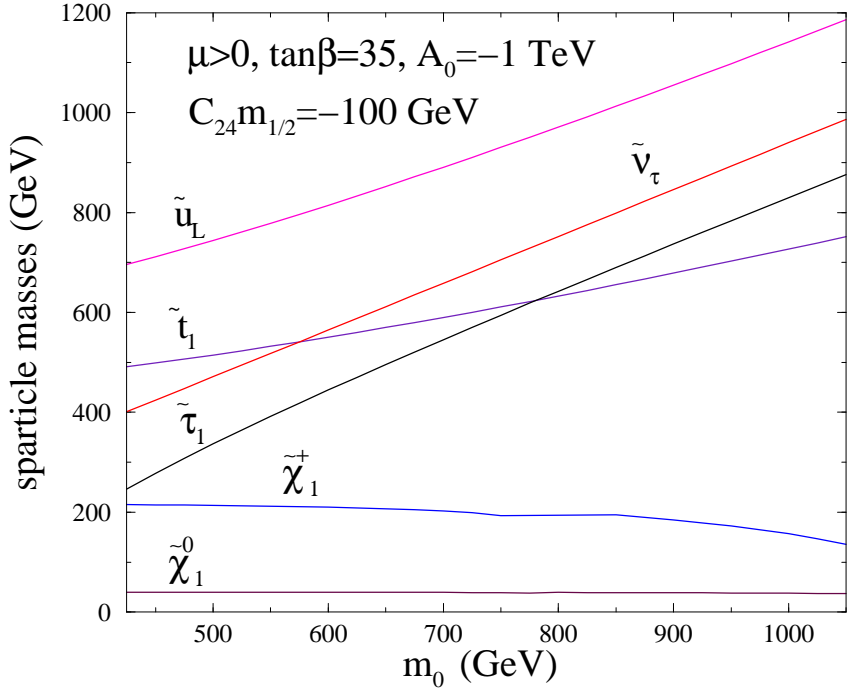}      
\end{minipage}}                       
\caption{
(a):Sparticle masses in mSUGRA vs $m_0$ for the case $\mu>0$,
$\tan\beta =35$, $A_0=0$, and $m_{\frac{1}{2}}=300$ GeV.
 (b): Sparticle masses vs $m_0$ in the 24 plet case of
nonuniversality. The value of $m_3(M_G)$ and other parameters are 
chosen to be identical with that of (a) for comparison.
(c):Similar to (b) except that $A_0=-1$ TeV
and $C_{24}m_{1/2}=-100$ GeV. The parameters chosen in Fig.(4b)
and Fig.(4c) favor 
$b-\tau$ unification as may be seen from Figs.~\ref{perc_params_a} to
\ref{perc_params_d}.
}                       
\label{unification_spectra}
\end{figure} 

%%%%%%%%%%%%%%%%%%%%%%%%%%%%%%%%%%%%%%%%%%%%%%%%%%%%%%%%

%%%%%%%%%%%%%%%%%%%%%%%%%%%%%%%%%%%%%%%%%%%%%%%%%%%%%%%%%%%%%%%%%%%%%%%%%%%%%
\newpage
\begin{figure}
\hspace*{-0.6in}
\centering
\includegraphics[width=\textwidth]{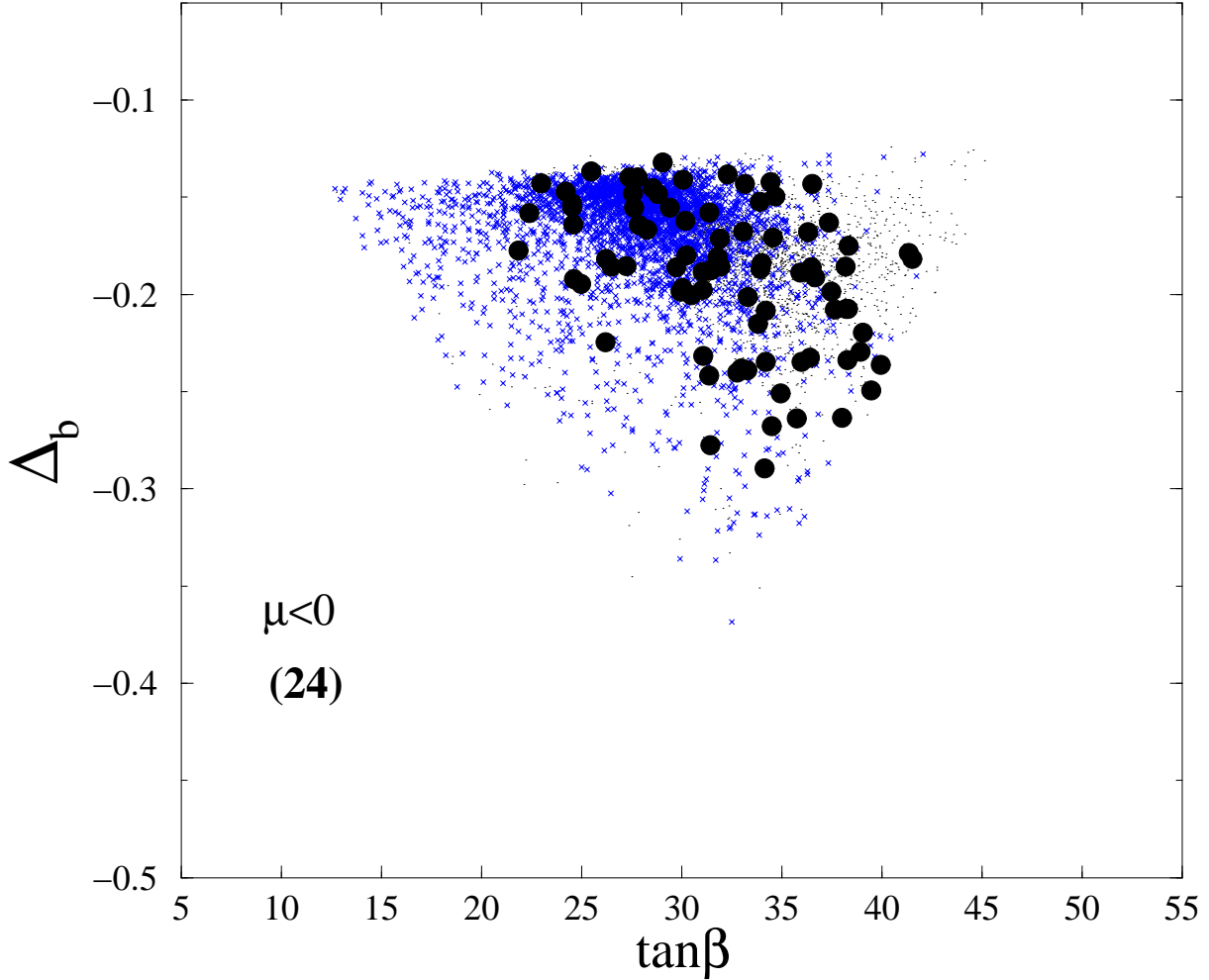}
\caption{The superymmetric correction  $\Delta_b$  to the  
b-quark mass  vs $\tan\beta$ for the 
24-plet case for $\mu<0$ when $\delta_{b \tau}<20$\%.  Here 
$0<m_0<2$ TeV, $-1 ~{\rm TeV} <C_{24}m_{1/2}<1 ~{\rm TeV}$,
$-6 ~{\rm TeV}<A_0<6~{\rm TeV}$.
%Additionally, third generation trilinear parameters are all 
%different, varying between 
%$-6 ~{\rm TeV}$ to $6~{\rm TeV}$.
 The dots refer to points satisfying $b-\tau$ 
unification with $\delta_{b \tau}<20$\%. 
The (blue) crosses additionally satisfy 
$b \rightarrow s+ \gamma$ limits. 
The (black) filled 
circles satisfy in addition the muon $g-2$ constraint
corresponding to the points in the parameter space where $m_2\mu>0$, 
typically with 10\%$<\delta_{b \tau}<20$\%. 
%%Thus the muon 
%%$g-2$ result disfavors the $\mu<0$ case of the 24-plet 
%%scenario compared to the $\mu>0$ branch.
}
\label{muneg_uni}
\end{figure} 

%%%%%%%%%%%%%%%%%%%%%%%%%%%%%%%%%%%%%%%%%%%%%%%%%%%%%%%%
%%%%%%%%%%%%%%%%%%%%%%%%%%%%%%%%%%%%%%%%%%%%%%%%%%%%%%%

\newpage
\begin{figure}                       
\vspace*{-2.0in}
\subfigure[]{
\label{bs_amu_region_a}                  
\hspace*{-0.6in}
\begin{minipage}[b]{\textwidth}                       
\centering
\includegraphics[width=0.65\textwidth]{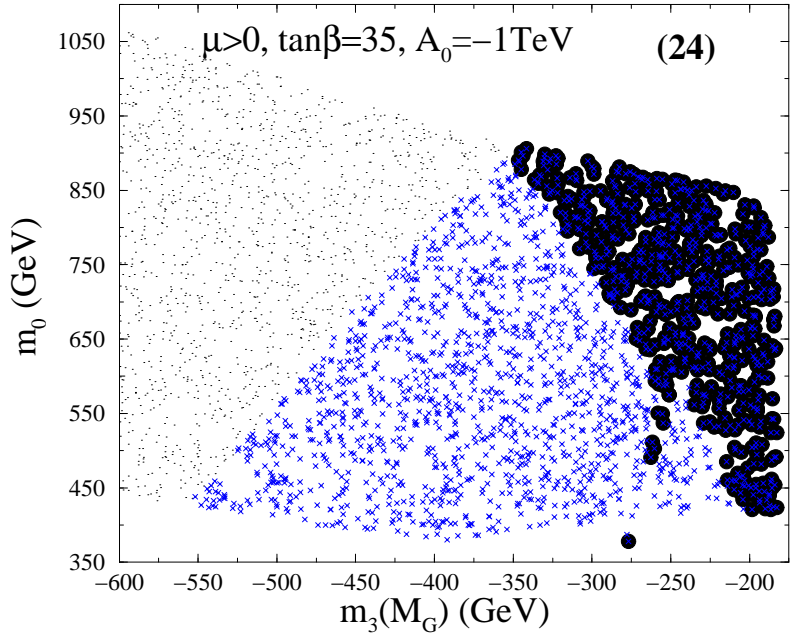}
\end{minipage}}                       

% 1 line gap given here for the next fig to be at the bottom
\subfigure[]{
\label{bs_amu_region_b}    
\hspace*{-0.6in}            
\begin{minipage}[b]{\textwidth}                       
\centering                      
\includegraphics[width=0.65\textwidth]{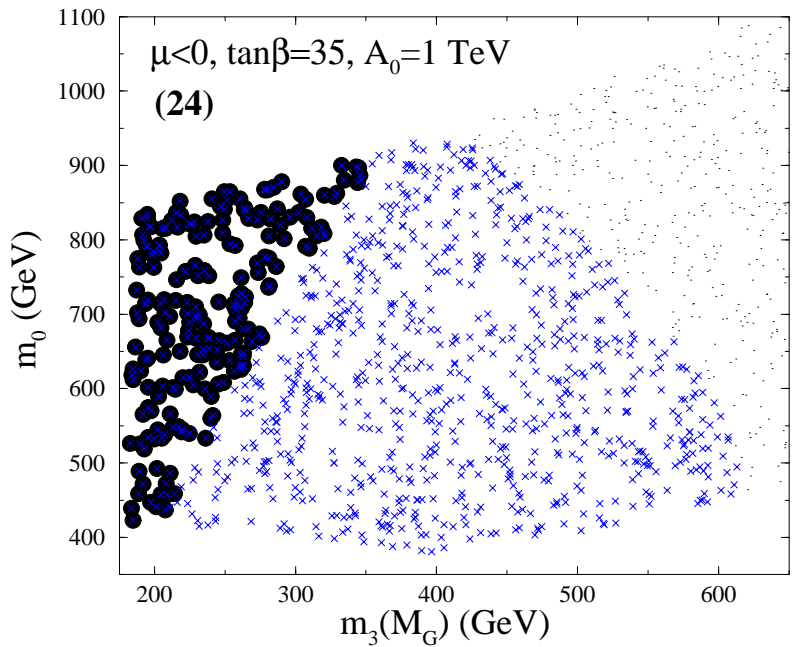}
\end{minipage}}                       
\caption{The allowed and the disallowed regions for the 24-plet case in the
$m_0-m_3(M_G)$ plane with various constraints. The dotted points satisfy
 $b-\tau$ unification with $\delta_{b-\tau}<0.3$ but do not satisfy
the $b\rightarrow s +\gamma$ constraint or the muon $g-2$ constraint.
Blue crosses are the allowed points which satisfy the 
$b-\tau$ unification with $\delta_{b \tau}<0.3$ and the muon $g-2$
constraint.
Black filled  circles are the allowed points which satisfy all the
constraints, i.e., the $b-\tau$ unification with $\delta_{b-\tau}<0.3$, 
 the muon $g-2$ constraint, and the  $b\rightarrow s +\gamma$ 
 constraint. (a) is for the case when $\mu>0$, $\tan\beta =35$,
 and $A_0=1$ TeV and (b) is for the case when 
 $\mu<0$, $\tan\beta =35$ and $A_0=-1$ TeV.}
\label{bs_amu_region}    
\end{figure} 
%%%%%%%%%%%%%%%%%%%%%%%%%%%%%%%%%%%%%%%%%%%%%%%%%%%%%%%%%%%%%%%%%%%%%%%%%%%%

%%%%%%%%%%%%%%%%%%%%%%%%%%%%%%%%%%%%%%%%%%%%%%%%%%%%%%%%
\newpage
\begin{figure}           
\vspace*{-2.0in}                                 
\subfigure[]{                       
\label{mhalf_gaugino_5_a} 
\hspace*{-0.6in}                     
\begin{minipage}[b]{0.5\textwidth}                       
\centering
\includegraphics[width=\textwidth,height=\textwidth]{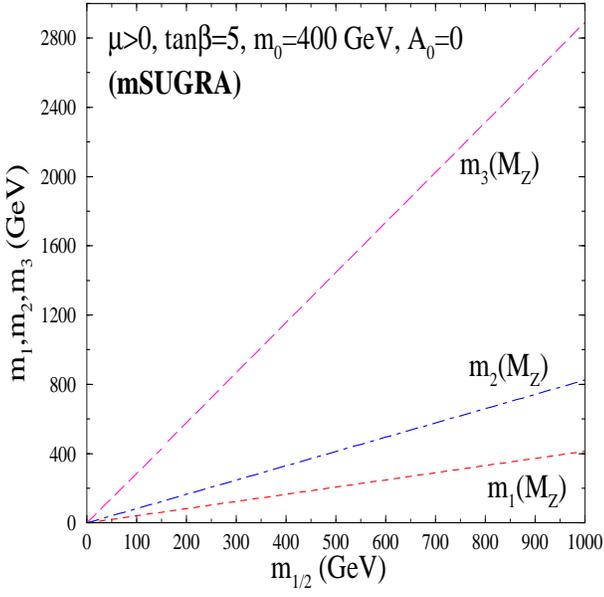}
\end{minipage}}                       
\hspace*{0.3in}
\subfigure[]{                       
\label{mhalf_gaugino_5_b}                     
\begin{minipage}[b]{0.5\textwidth}                       
\centering                      
\includegraphics[width=\textwidth,height=\textwidth]{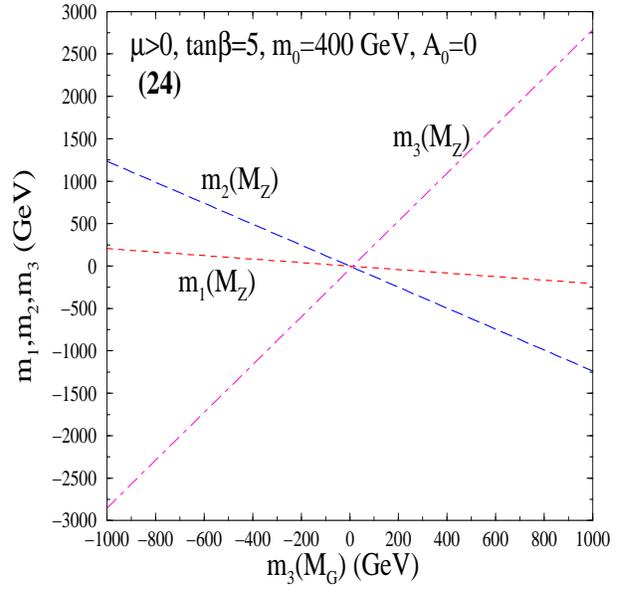}
\end{minipage}}                       
\hspace*{-0.6in}                     
\subfigure[]{                       
\label{mhalf_gaugino_5_c}  
\begin{minipage}[b]{0.5\textwidth}                       
\centering
\includegraphics[width=\textwidth,height=\textwidth]{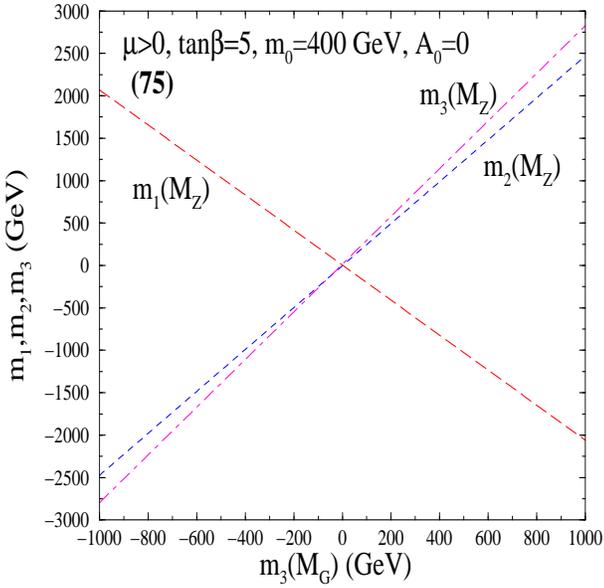}
\end{minipage}}
\hspace*{0.3in}                       
\subfigure[]{                       
\label{mhalf_gaugino_5_d} 
\begin{minipage}[b]{0.5\textwidth}                       
\centering                      
\includegraphics[width=\textwidth,height=\textwidth]{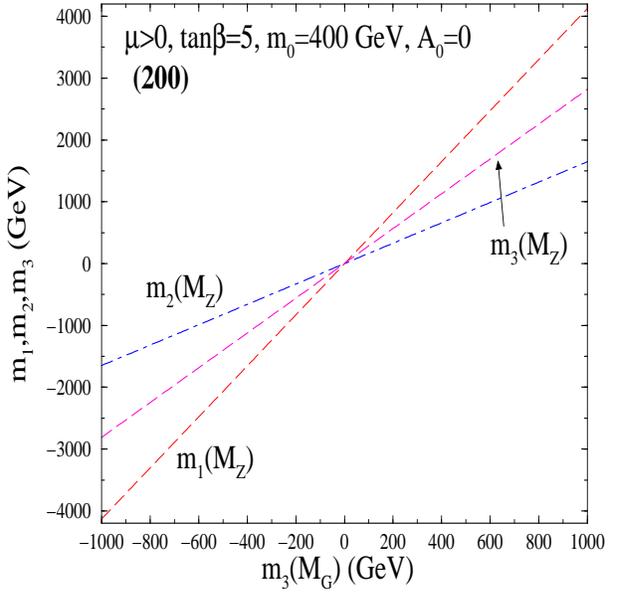}
\end{minipage}}                       
\caption{
(a):Gaugino masses in mSUGRA vs $m_{1/2}$ when $\mu>0$, $\tan\beta=5$,
$m_0=400$ GeV, and $A_0=0$.
(b): gaugino masses for the 24 plet nonuniversal case vs $m_3(M_G)$
for the same set of parameters as in (a).
(c): Same as (b) except for the 75-plet case. (d): Same as (b) 
except for the 200-plet case.  }                       
\label{mhalf_gaugino_5}
\end{figure}

%%%%%%%%%%%%%%%%%%%%%%%%%%%%%%%%%%%%%%%%%%%%%%%%%%%%%%%%%

\newpage
\begin{figure}           
\vspace*{-2.0in}                                 
\subfigure[]{                       
\label{vary_m0_5_a} 
\hspace*{-0.6in}                     
\begin{minipage}[b]{0.5\textwidth}                       
\centering
\includegraphics[width=\textwidth,height=\textwidth]{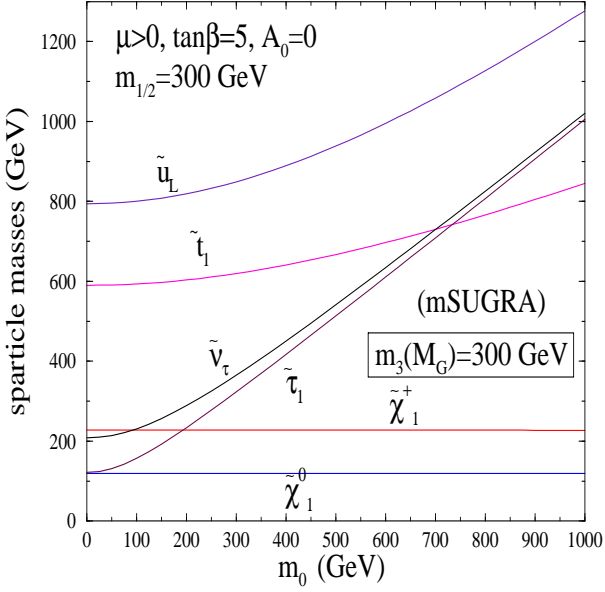}    
\end{minipage}}                       
\hspace*{0.3in}
\subfigure[]{                       
\label{vary_m0_5_b}                         
\begin{minipage}[b]{0.5\textwidth}                       
\centering                      
\includegraphics[width=\textwidth,height=\textwidth]{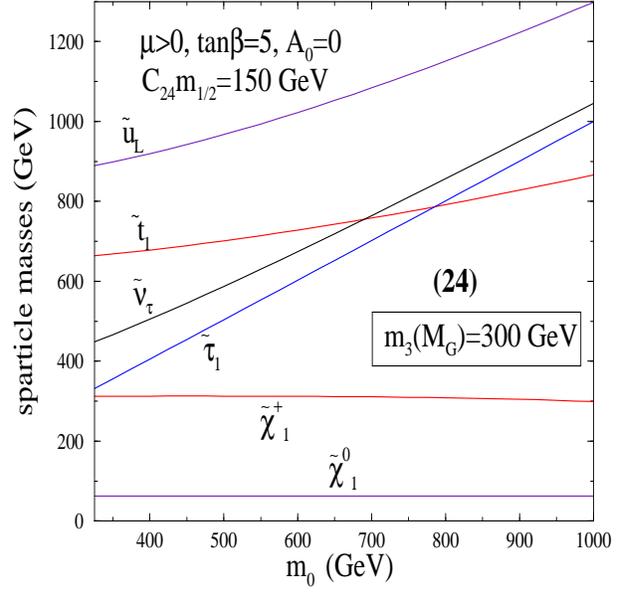} 
\end{minipage}}                       
\hspace*{-0.6in}                     
\subfigure[]{                       
\label{vary_m0_5_c}                      
\begin{minipage}[b]{0.5\textwidth}                       
\centering
\includegraphics[width=\textwidth,height=\textwidth]{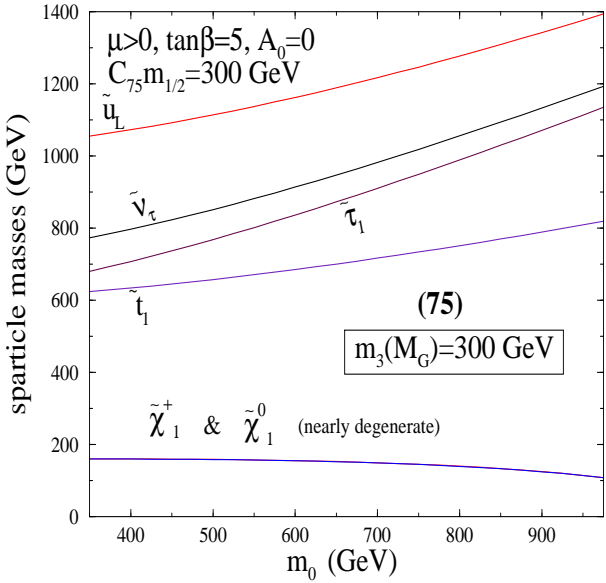}
\end{minipage}}
\hspace*{0.3in}                       
\subfigure[]{                       
\label{vary_m0_5_d} 
\begin{minipage}[b]{0.5\textwidth}                       
\centering                      
\includegraphics[width=\textwidth,height=\textwidth]{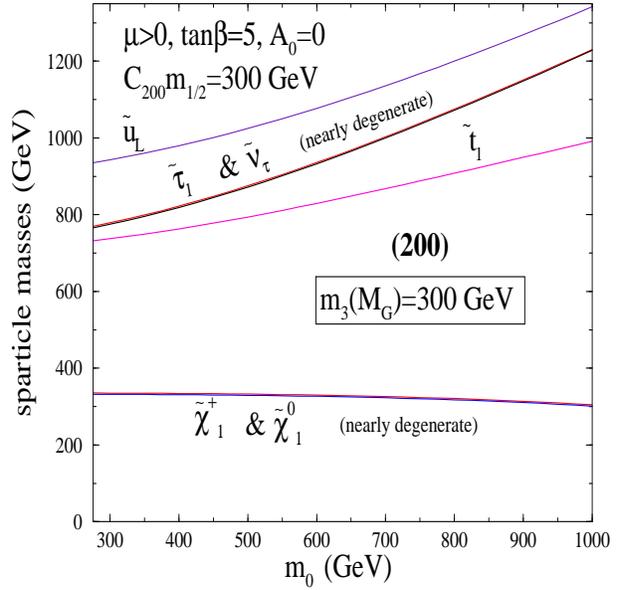}
\end{minipage}}                       
\caption{
(a):Sparticle masses in mSUGRA vs $m_0$  
for $\mu>0$ when $\tan\beta =5$, $A_0=0$, and $m_{\frac{1}{2}}=300$ GeV. 
(b): Sparticle masses for the 24-plet nonuniversal case 
for $\mu>0$, when $\tan\beta =5$, $A_0=0$, and $m_3(M_G)=300$ GeV.
Values of $m_3(M_G)$ and of other parameters are 
chosen to be identical with that of (a) for a direct comparison.
(c): Same as (b) except for the 75-plet nonuniversal case. (d):
 Same as (b) except for the 200-plet nonuniversal case.  
}                       
\label{vary_m0_5} 
\end{figure}

%%%%%%%%%%%%%%%%%%%%%%%%%%%%%%%%%%%%%%%%%%%%%%%%%%%%%%

%%%%%%%%%%%%%%%%%%%%%%%%%%%%%%%%%%%%%%%%%%%%%%%%%%%%%%%%
\newpage
\begin{figure}           
\vspace*{-2.0in}                                 
\subfigure[]{                       
\label{vary_mhalf_5_a} 
\hspace*{-0.6in}                     
\begin{minipage}[b]{0.5\textwidth}                       
\centering
\includegraphics[width=\textwidth,height=\textwidth]{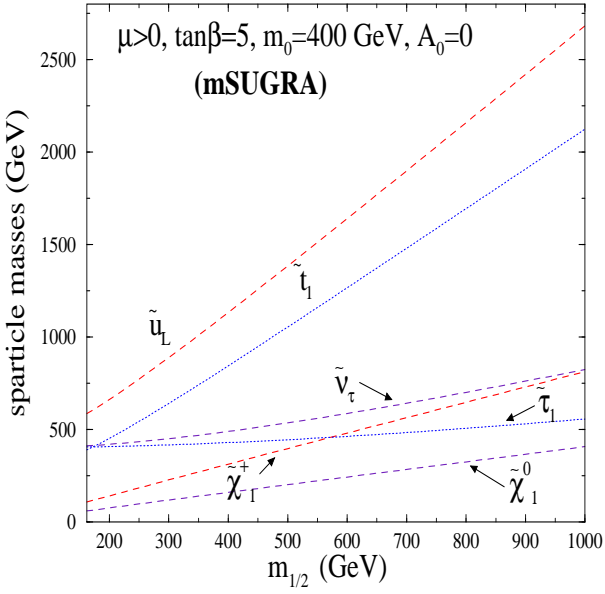}    
\end{minipage}}                       
\hspace*{0.3in}
\subfigure[]{                       
\label{vary_mhalf_5_b}                      
\begin{minipage}[b]{0.5\textwidth}                       
\centering                      
\includegraphics[width=\textwidth,height=\textwidth]{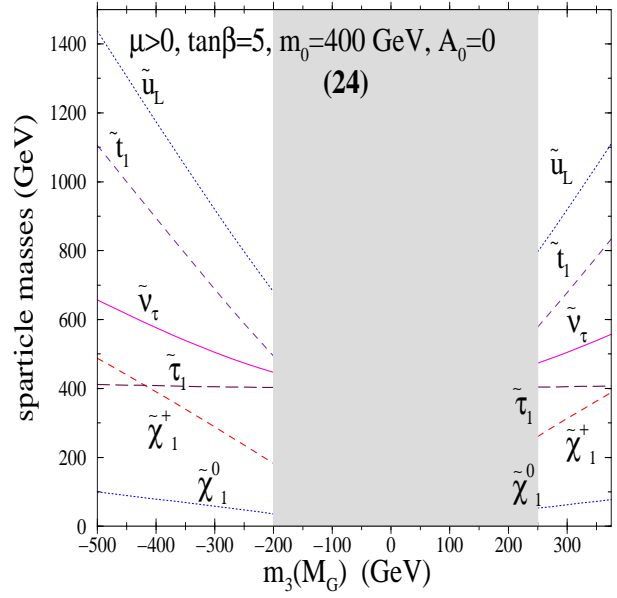} 
\end{minipage}}                       
\hspace*{-0.6in}                     
\subfigure[]{                       
\label{vary_mhalf_5_c}                        
\begin{minipage}[b]{0.5\textwidth}                       
\centering
\includegraphics[width=\textwidth,height=\textwidth]{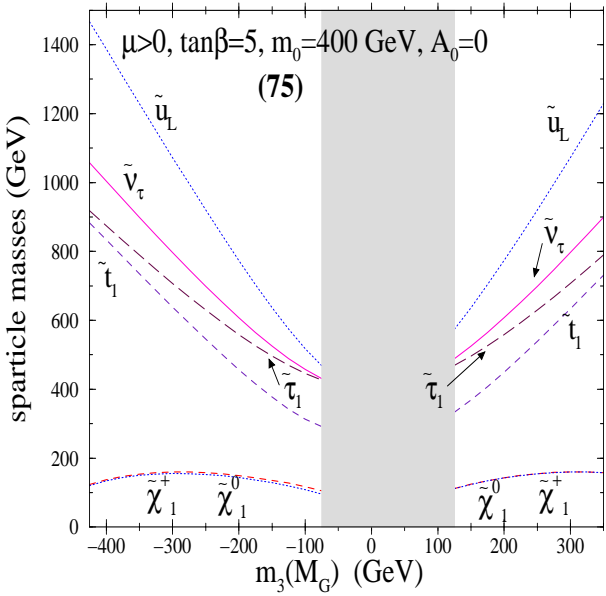}
\end{minipage}}
\hspace*{0.3in}                       
\subfigure[]{                       
\label{vary_mhalf_5_d} 
\begin{minipage}[b]{0.5\textwidth}                       
\centering                      
\includegraphics[width=\textwidth,height=\textwidth]{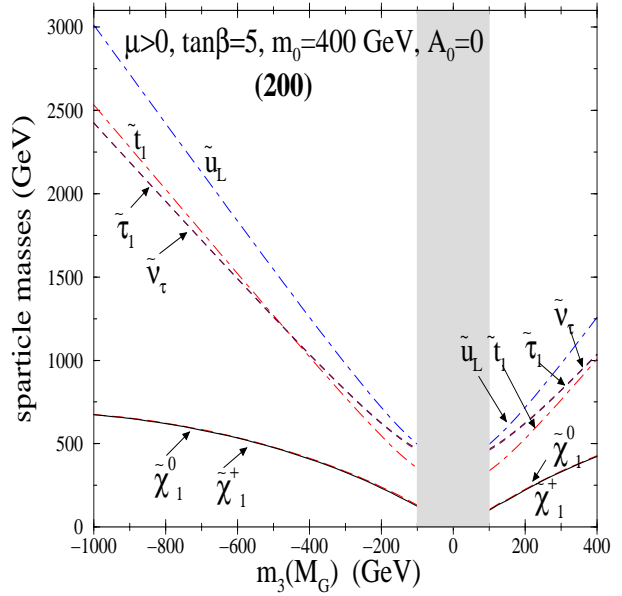}
\end{minipage}}                       
\caption{
(a):Sparticle masses in mSUGRA vs $m_{1/2}$ for the case $\mu>0$,
$\tan\beta =5$, $m_0=400$ GeV, and $A_0=0$.
(b): Similar plots of mass spectra vs $m_3(M_G)$ for the 24-plet 
nonuniversal case. (c): Same as (b) except for the 75-plet nonuniveral case.
 (d): Same as (b) except for the 200-plet nonuniversal case. 
 The gray areas represent the disallowed regions because 
of chargino mass limits and the absence of 
radiative electroweak symmetry breaking.}                       
\label{vary_mhalf_5}
\end{figure}

%%%%%%%%%%%%%%%%%%%%%%%%%%%%%%%%%%%%%%%%%%%%%%%%%%%%%%%%
\newpage
\begin{figure}           
\vspace*{-2.0in}                                 
\subfigure[]{                       
\label{other_perc_params_a} 
\hspace*{-0.6in}                     
\begin{minipage}[b]{0.5\textwidth}                       
\centering
\includegraphics[width=\textwidth,height=\textwidth]{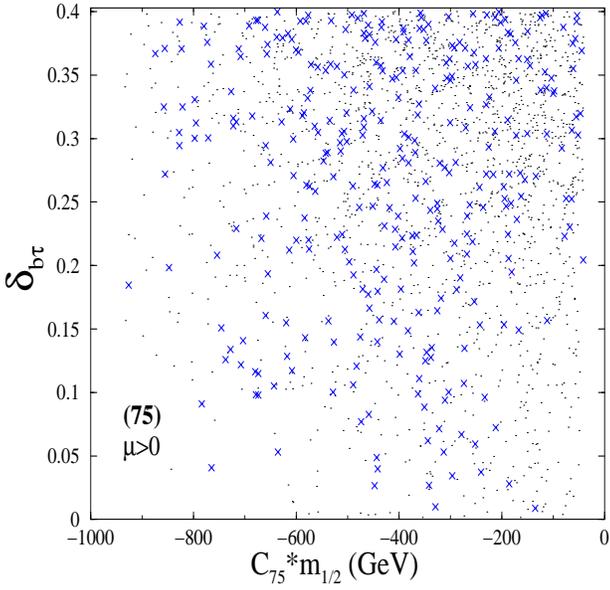}    
\end{minipage}}                       
\hspace*{0.3in}
\subfigure[]{                       
\label{other_perc_params_b}                       
\begin{minipage}[b]{0.5\textwidth}                       
\centering                      
\includegraphics[width=\textwidth,height=\textwidth]{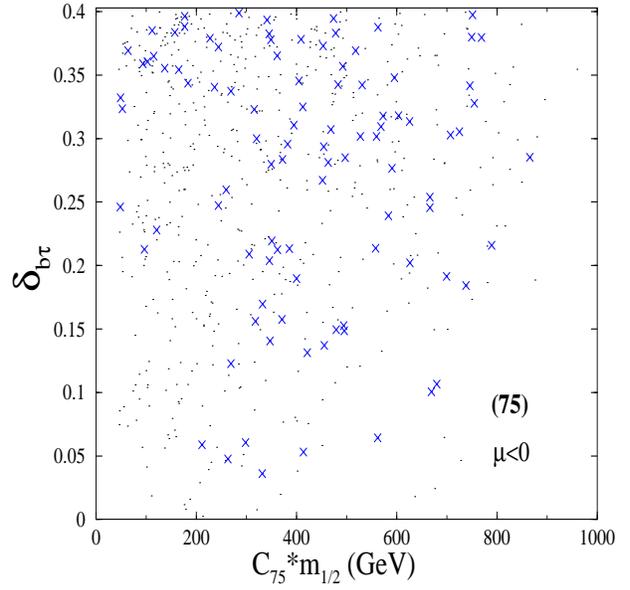}
\end{minipage}}                       
\hspace*{-0.6in}                     
\subfigure[]{                       
\label{other_perc_params_c}                      
\begin{minipage}[b]{0.5\textwidth}                       
\centering
\includegraphics[width=\textwidth,height=\textwidth]{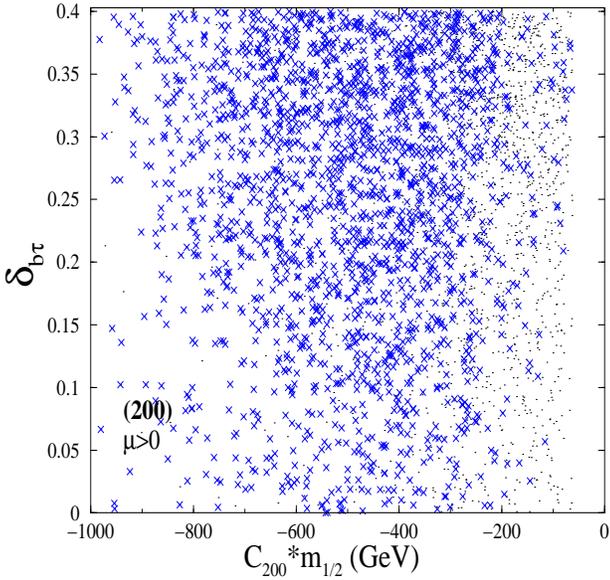}
\end{minipage}}
\hspace*{0.3in}                       
\subfigure[]{                       
\label{other_perc_params_d}                       
\begin{minipage}[b]{0.5\textwidth}                       
\centering                      
\includegraphics[width=\textwidth,height=\textwidth]{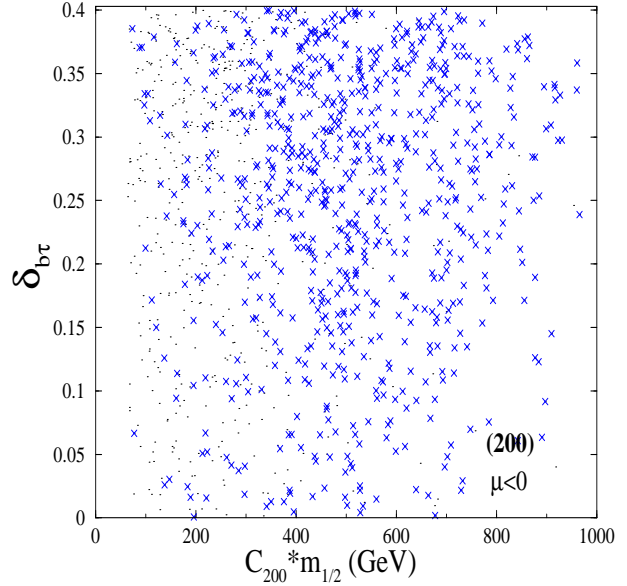}
\end{minipage}}                       
\caption{(a): Plot of $\lambda_b-\lambda_{\tau}$ unification parameter 
$\delta_{b \tau}$ vs $C_{75}m_{1/2}$ for the 75-plet case 
when $\mu>0$ and the other parameters are varied so that 
 $\tan\beta<55$, $0<m_0<2$ TeV, and the third generation trilinear parameters
 are varied independently in the range $-6 ~{\rm TeV}$ to $6~{\rm TeV}$.
The dotted points satisfy $b-\tau$ unification at the level shown and
the (blue) crosses additionally obey the 
$b \rightarrow s+ \gamma$ constraint.  There are no parameter points 
consistent with the muon $g-2$ constraint.
(b) Same as (a)except that $\mu<0$.
(c) Same as (a) except that the plot is for $\delta_{b \tau}$ vs 
$C_{200}m_{1/2}$. (d) Same as (c) exept that $\mu<0$.
}                       
\label{other_perc_params} 
\end{figure} 
%%%%%%%%%%%%%%%%%%%%%%%%%%%%%%%%%%%%%%%%%%%%%%%%%%%%%%%%

\end{document}